\documentclass[preprint,aps,showpacs]{revtex4}
\usepackage[dvips]{graphicx}
\newcommand{\bp} {\mbox{\boldmath$p$}}
\newcommand{\bq} {\mbox{\boldmath$q$}}
\newcommand{\bj} {\mbox{\boldmath$j$}}
\newcommand{\bL} {\mbox{\boldmath$L$}}
\newcommand{\bsigma} {\mbox{\boldmath$\sigma$}}
\newcommand{\btheta} {\mbox{\boldmath$\theta$}}

\begin{document}

\title{Second moment of the Husimi distribution as a measure of
complexity of quantum states}
\author{Ayumu Sugita\thanks{sugita@rcnp.osaka-u.ac.jp}}

\affiliation{Research Center for Nuclear Physics, Osaka University,\\
10-1 Mihogaoka, Ibaraki, Osaka 567-0047, Japan}

\author{Hirokazu Aiba\thanks{aiba@koka.ac.jp}}

\affiliation{Kyoto Koka Women's College,\\
38 Kadono-cho Nishikyogoku, Ukyo-ku, 615-0882, Japan}

\date{\today}

\begin{abstract}
We propose the second moment of the Husimi distribution as a measure
of complexity of quantum states. The inverse of this quantity
represents the effective volume in phase space occupied by the Husimi 
distribution, and has a good correspondence with chaoticity of 
classical system. Its properties are similar to the classical entropy
proposed by Wehrl, but it is much easier to calculate
numerically. 
We calculate this quantity in the quartic oscillator model, and show that
it works well as a measure of chaoticity of quantum states.
\end{abstract}

\maketitle


\newpage

\section{Introduction}

Although the quantum manifestation of chaos has been extensively studied
over the past few decades, to define ``quantum chaos'' is still the
main problem in this field. The direct extension of the definition 
applicable to
classical chaos seems to fail because of the linearity of the
Schr\"{o}dinger equation. Since quantum mechanics contains
classical mechanics as a limit, however, 
there must be something in quantum mechanics that produces
classical chaos in the classical limit.

There have been many attempts to define a measure of
quantum chaos. 
There are some measures using the level statistics, but 
much more information could be obtained from analysises 
of individual quantum states. We can classify measures of complexity of 
quantum states into two types: 
\begin{enumerate}
\item complexity of pure states
\item complexity of an ensemble of quantum states
\end{enumerate}
For example, the von Neumann entropy  
of the density matrix belongs to type (2).
In this paper, we focus on complexity of type (1). 

Some quantities defined in terms of the expansion coefficients 
are often used in numerical calculations.
For instance, suppose a quantum state is expanded in
an appropriate basis $\left\{|i\rangle\right\}$:
\begin{eqnarray}
|\varphi\rangle = \sum_{i}c_{i}|i\rangle.
\end{eqnarray}
Let us define $p_{i}=|c_{i}|^{2}$. Then the 
information (Shannon) entropy
\begin{eqnarray}
S = - \sum_{i}p_{i}\ln p_{i},
\end{eqnarray}
and moments of the distribution 
\begin{eqnarray}
M_{k} = \sum_{i}p_{i}^{k},
\end{eqnarray}
are measures of localization (or
delocalization) with respect to this basis.  
In particular, the inverse of the second moment $M_{2}^{-1}$
is called the 
number of principal components (NPC): it becomes unity
when the state has only one component, while it becomes $n$
when the probability is equally distributed over $n$ basis vectors. 
Such quantities are easy to calculate. However, an obvious defect is that 
the definition depends on the basis.

Wehrl has proposed a good measure of complexity of quantum states
based on the Husimi distribution function $\rho_{H}(\bp,\bq)$
\cite{wehrl}\cite{wehrl2}. 
He called it ``classical entropy'', 
\begin{eqnarray}
S(\rho_{H}) = \int d\bp d\bq\;\rho_{H}\ln\rho_{H},
\end{eqnarray}
while in his paper ``quantum entropy'' denotes 
the von Neumann entropy of a density matrix.   
Note that the classical entropy is applicable and
takes various values also for pure states,
while the quantum entropy always vanishes for them. 
Although Wehrl introduced the classical entropy as an approximation
to the quantum entropy, their values are not necessarily close
even in the limit $\hbar\rightarrow 0$. (See the discussion in 
\cite{takahashi}.)
According to our
classification, we discuss the classical entropy to describe
complexity of type (1), i.e. of pure states.

Chaoticity in classical mechanics can be characterized by 
the delocalization of orbits. In integrable systems, there are many
constraints from symmetries, such that orbits are confined to
low-dimensional tori. As the system becomes chaotic, the tori
are destroyed and orbits can spread to higher dimensional
space. In highly chaotic systems, orbits spread uniformly 
over the equi-energy surface. 
Such systems are called ergodic. 
Among the several conditions to characterize chaoticity of classical
mechanics, the ergodicity is a rather weak one. For example, any 
one-dimensional time-independent Hamiltonian
system is ergodic, but never chaotic. 
However, in physically natural situations of many-dimensional systems,
the ergodicity works as a definition of classical chaos.

We can expect a similar behavior in quantum mechanics.
The Husimi function is a function on phase space, and takes only
non-negative values while the Wigner function can be negative and 
is usually violently oscillating \cite{takahashi}.
Hence the Husimi function can be regarded as a probability distribution 
in phase 
space, and its order of delocalization can be a measure
of chaoticity of quantum states. In that sense, Wehrl's classical 
entropy seems to be a good candidate measure. 

However, Wehrl's entropy is defined as an integral 
over phase space, and is not easy to calculate numerically.
To the best of the authors' knowledge, 
there is no calculation for a more than
one-dimensional system. (A calculation for a one-dimensional
time-driven system is in \cite{takahashi}.) 
The difficulty mainly comes from the redundancy of 
the Husimi function: the Husimi function of $k$ dimensional
system is a function on
$2k$ dimensional space, while 
a function on $k$ dimensional space is enough
to keep the same information in either coordinate
or momentum representation.

Complex pure quantum states are usually represented as sets of 
expansion coefficients with respect to a basis. This will be the case
also in our following discussion.
When we calculate Wehrl's entropy of a quantum state, the normal procedure
is the following:
\begin{enumerate}
\item Calculate the Husimi function $\rho_{H}$ on many sampling
points $\left\{(\bp_{i},\bq_{i})\right\}$.
\item Take an average $\langle\rho_{H}\ln\rho_{H}\rangle$ over 
the sampling points.
\end{enumerate}
However, this 
procedure seems to be overly excessive 
because we have all the information about
the quantum state in the set of expansion coefficients. 
There must exists some formula to calculate the average directly from
the coefficients.
The main concern of this paper is to avoid the redundancy of
the Husimi function in numerical evaluation of a suitable
measure of complexity. 
In other words, we wish to know how to get some
averages related to the Husimi function without 
actually calculating the Husimi function itself.   

It seems difficult to derive a simple formula for the entropy because
of the transcendental logarithmic function. 
However, a simple formula is possible for an algebraic function. 
The second moment, which is the average
of the square of the distribution function, is especially easy to
calculate. We will show later a formula in which  the 
second moment of the Husimi distribution is expressed directly
in terms of expansion coefficients in the harmonic oscillator basis.
(If the quantum states are given by expansion coefficients in 
another basis, we should calculate the transformation matrix and
change the basis to the harmonic oscillator basis.) 
The numerical effort is of order
$N^{2}$ for a quantum state where $N$
is the number of basis vectors. If we calculate the second moments
for all quantum states given by diagonalization of a $N\times N$
matrix, the numerical effort is of order $N^{3}$, which is the same
as the order of the diagonalization of the matrix. 

We represent the inverse of the second moment by $W_{2}$:
\begin{eqnarray}
W_{2}(\rho_{H}) = \frac{1}{M_{2}(\rho_{H})},
\end{eqnarray}
\begin{eqnarray}
M_{2}(\rho_{H}) = 
\int\frac{d\bp d\bq}{(2\pi\hbar)^{k}}\; \rho_{H}(\bp,\bq)^{2}.
\end{eqnarray}
$W_{2}$ represents the effective phase space volume occupied by
the Husimi function.
For example, if $\rho_{H}$ takes the same value
over a region with volume $V$ and takes zero value outside of it, $W_{2}=V$. 
The unit of $W_{2}$ is the Planck cell volume.

Next we summarize the main points of this paper. We use the second
moment of the Husimi distribution to define complexity
of quantum states. It is a measure of delocalization of the Husimi
distribution in phase space.
It is defined in a base-independent way and has a good correspondence 
with complexity of classical mechanics. Moreover, we can calculate
it directly from expansion coefficients, without calculating the redundant
Husimi function. Therefore it is not so 
difficult to evaluate numerically.

This paper is organized as follows: In section \ref{husimi}
we summarize the definition and some properties of the Husimi distribution
function. In section \ref{second} we derive the formula to calculate
the second moment of the Husimi distribution directly from expansion 
coefficients in the harmonic oscillator basis. In section \ref{numerical}
we introduce a model Hamiltonian and show numerical results 
which illustrate the meaning of $W_{2}$ as a measure of complexity. 
The final section is devoted to a summary. In appendix we show details
of semiclassical calculations of the second moment of the Husimi 
distribution in integrable and ergodic limits.

\section{Husimi distribution function}\label{husimi}
In this section, we review some properties of the
Husimi function \cite{husimi}. We restrict ourselves to a 
one-dimensional system
for simplicity, but the generalization to many-dimensional systems
is straightforward. 

The Husimi function of a quantum state $|\varphi\rangle$ is 
defined as 
\begin{eqnarray}
\rho_{H,\lambda}(p,q) = |\langle z,\lambda|\varphi\rangle |^{2}.
\label{def}
\end{eqnarray}
Here, $|z,\lambda\rangle$ is a coherent state defined 
as an eigenstate for a complex eigenvalue $z$
\begin{eqnarray}
\hat{a}_{\lambda}|z,\lambda\rangle = z|z,\lambda\rangle,
\end{eqnarray}
where $\hat{a}_{\lambda}$ is an operator 
with $\lambda$ as an arbitrary parameter
\begin{eqnarray}
\hat{a}_{\lambda} = \frac{1}{\sqrt{2\hbar}}
\left(\sqrt{\lambda}\hat{q} + i\frac{\hat{p}}{\sqrt{\lambda}}\right).
\end{eqnarray}
The real and imaginary parts of $z$ are related to the phase space point 
$(p,q)$ by
\begin{eqnarray}
q & = & \sqrt{\frac{\hbar}{2\lambda}}(z+\bar{z}), \\
p & = & -i\sqrt{\frac{\lambda\hbar}{2}}(z-\bar{z}).
\end{eqnarray}

It is not difficult to derive the relation
\begin{eqnarray}
\rho_{H,\lambda}(p,q) & = & 
\frac{1}{\pi\hbar}\int dp'dq'\rho_{W}(p',q')\nonumber \\
&& \times
\exp\left[-\frac{1}{\hbar}\left\{\lambda (q'-q)^{2} +
\frac{(p'-p)^{2}}{\lambda}\right\}\right],
\end{eqnarray}
where $\rho_{W}$ is the Wigner function
\begin{eqnarray}
\rho_{W}(p,q) = \int d\eta 
\langle q - \eta /2 | \varphi \rangle 
\langle \varphi | q + \eta /2 \rangle
e^{ip\eta/\hbar}.
\end{eqnarray}
A state $|q\rangle$ represents an eigenstate of the coordinate
representation.
There we see that the Husimi function is 
a kind of the coarse-grained Wigner function.
The main advantage of the Husimi function is that it is 
non-negative, as is obvious from the definition (\ref{def}).
Hence the Husimi function can be formally 
regarded as a probability distribution,
whereas the Wigner function can not.

\section{Second moment of the Husimi distribution}\label{second}
In this section, we consider the second moment of the Husimi distribution
\begin{eqnarray}
M_{2}(\rho_{H}) & = & \int \frac{dpdq}{2\pi\hbar}\rho_{H}(p,q)^{2},\\
& = & \int \frac{d^{2}z}{\pi}|\langle z|\varphi\rangle|^{4}.
\end{eqnarray}
Here and in the following we omit $\lambda$ for simplicity. 

Suppose the quantum state $|\varphi\rangle$ is represented 
in the harmonic 
oscillator basis
\begin{eqnarray}
|\varphi\rangle & = & \sum_{n=0}^{\infty}c_{n}|n\rangle,
\label{expand}
\end{eqnarray}
then $M_{2}$ can be expanded as
\begin{eqnarray}
M_{2}(\rho_{H}) & = & 
\sum_{n,n'}\sum_{m,m'} 
\frac{c_{n}c_{n'}c_{m}^{*}c_{m}^{*}}{\sqrt{n!n'!m!m'!}}
\int \frac{d^{2}z}{\pi} e^{-2|z|^{2}}\bar{z}^{n+n'}z^{m+m'}, \\
& = &
\sum_{n,n'}\sum_{m,m'}
\frac{c_{n}c_{n'}c_{m}^{*}c_{m}^{*}}{\sqrt{n!n'!m!m'!}}
\frac{(n+n')!}{2^{n+n'+1}}\delta_{n+n',m+m'}. \\
\end{eqnarray}
Here, we used the following formulas
\begin{eqnarray}
\langle n|z\rangle = e^{-|z|^{2}/2}\frac{z^{n}}{\sqrt{n!}},
\end{eqnarray}
and
\begin{eqnarray}
\int \frac{d^{2}z}{\pi}e^{-|z|^{2}}\bar{z}^{n}z^{m} =
n!\delta_{n,m}.
\end{eqnarray}
As a result, we obtain the formula
\begin{eqnarray}
M_{2}(\rho_{H}) & = &
\frac{1}{2}\sum_{L=0}^{\infty}|B_{L}|^{2},
\label{sum_m2}
\end{eqnarray}
where
\begin{eqnarray}
B_{L} & = & \sum_{j=0}^{L}
\sqrt{\frac{L!}{2^{L}j!(L-j)!}}\; c_{j}c_{L-j}.
\end{eqnarray}
Since there is a truncation $n\le N$ in the expansion 
(\ref{expand}) in any real calculation, the sum (\ref{sum_m2})
is also finite. The numerical effort is of order $N^{2}$
for a quantum state.

The generalization to many-dimensional systems is straightforward.
The result is
\begin{eqnarray}
M_{2}(\rho_{H}) & = & 
2^{-k}\sum_{\bL}\left|B_{\bL}\right|^{2},
\label{general1}
\end{eqnarray}
where $k$ is the dimension of the system and 
\begin{eqnarray}
B_{\bL} & = & \sum_{\bj\le\bL}
\sqrt{\frac{\bL!}{2^{\left|\bL\right|}\bj !\left(\bL-\bj\right)!}}\;
c_{\bj}c_{\bL-\bj}.
\label{general2}
\end{eqnarray}
Here, $\bL$ and $\bj$ are $k$-dimensional vectors whose components are
non-negative integers. The factorial of those vectors means the
product of all factorials of the vector components. For example,
\begin{eqnarray}
\bL ! = \prod_{i=1}^{k}L_{i}!.
\end{eqnarray}
The absolute value denotes
\begin{eqnarray}
\left|\bL\right| = \sum_{i=1}^{k}L_{i},
\end{eqnarray}
and $\bj\le\bL$ means $j_{i}\le L_{i}$ for $\forall i$.

\section{Numerical results}\label{numerical}

\subsection{Model}

The model Hamiltonian we are considering here is 
that of a two-dimensional quartic oscillator
\begin{eqnarray}
H = \frac{1}{2}(p_{x}^{2} + p_{y}^{2}) + 
\frac{1}{2}(x^{4}+y^{4}) - kx^{2}y^{2},
\label{hamiltonian}
\end{eqnarray}
which has been adopted by many authors for the studies of level
statistics and wave functions \cite{bohigas}. We put $\hbar=1$
and regard all quantities as dimensionless.

This model has the simple 
scaling property, 
\begin{eqnarray}
H(\alpha^{2}p_{x},\alpha^{2}p_{y},\alpha x,\alpha y)
& = & \alpha^{4} H(p_{x},p_{y},x,y), 
\end{eqnarray} 
which means that the energy of the 
system does not essentially change the dynamics. 
However, the parameter $k$ changes the nature of the
system. The system is separable and integrable at $k=0$, and becomes
chaotic as $k$ increases. It is unbound for $k>1$.
Meyer \cite{meyer} has shown that for large
$k$ values ($\ge 0.4$) the classical phase space structure is almost
completely chaotic.  

This system has a discrete symmetry called $C_{4v}$: The Hamiltonian is 
invariant with respect to reflections about $x$-axis, $y$-axis and
also about the line $x=y$. In this paper we treat only 
quantum states which are symmetric under all reflections.
(This symmetry class is labeled $A_{1}$ in \cite{eckhardt}.)

\subsection{Structure of the classical phase space}

\begin{figure}
\begin{center}
\begin{minipage}{8cm}
\includegraphics[height=1.0\textwidth]{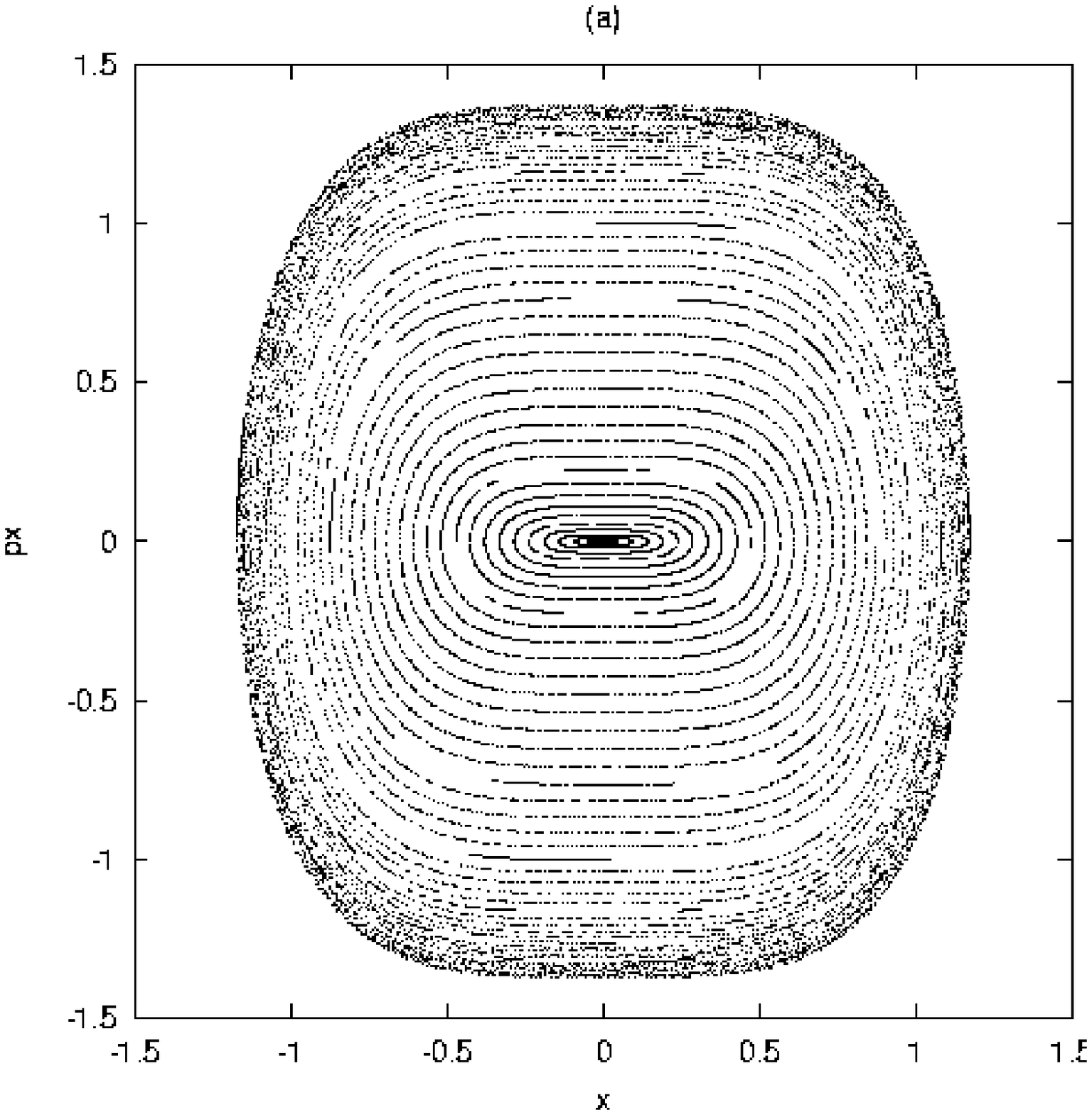}
\end{minipage}
\begin{minipage}{8cm}
\includegraphics[height=1.0\textwidth]{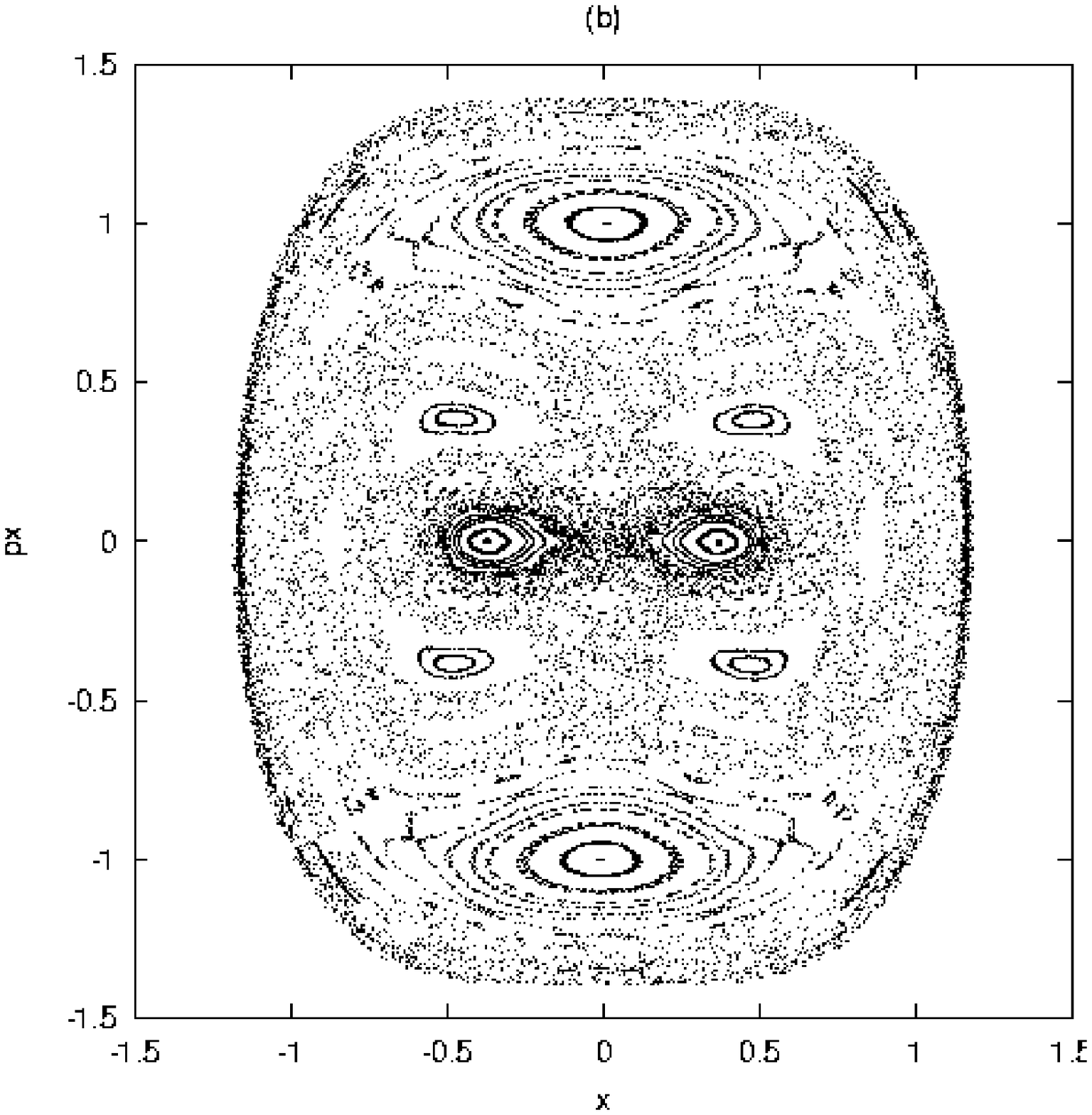}
\end{minipage}

\begin{minipage}{8cm}
\includegraphics[height=1.0\textwidth]{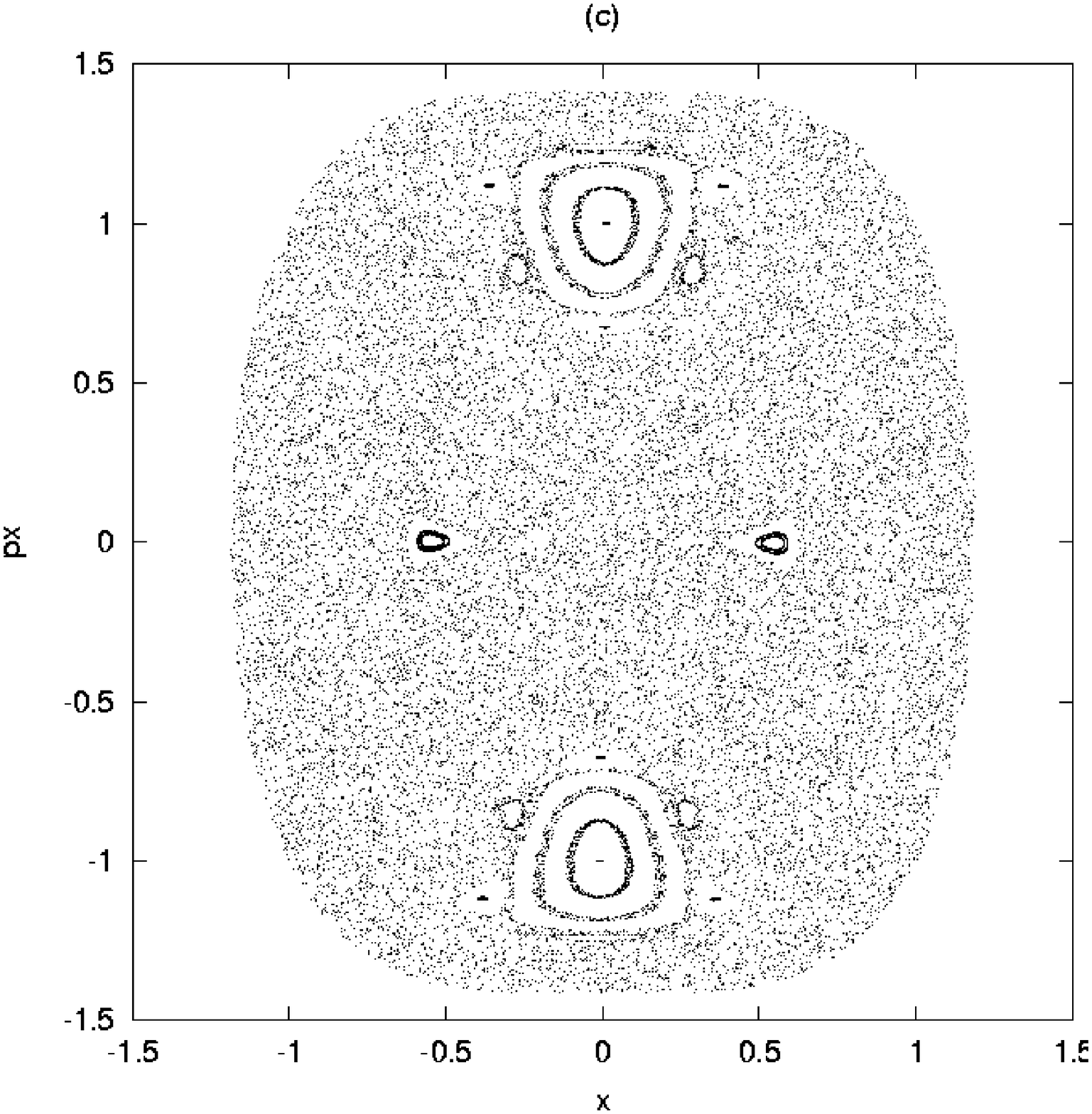}
\end{minipage}
\begin{minipage}{8cm}
\includegraphics[height=1.0\textwidth]{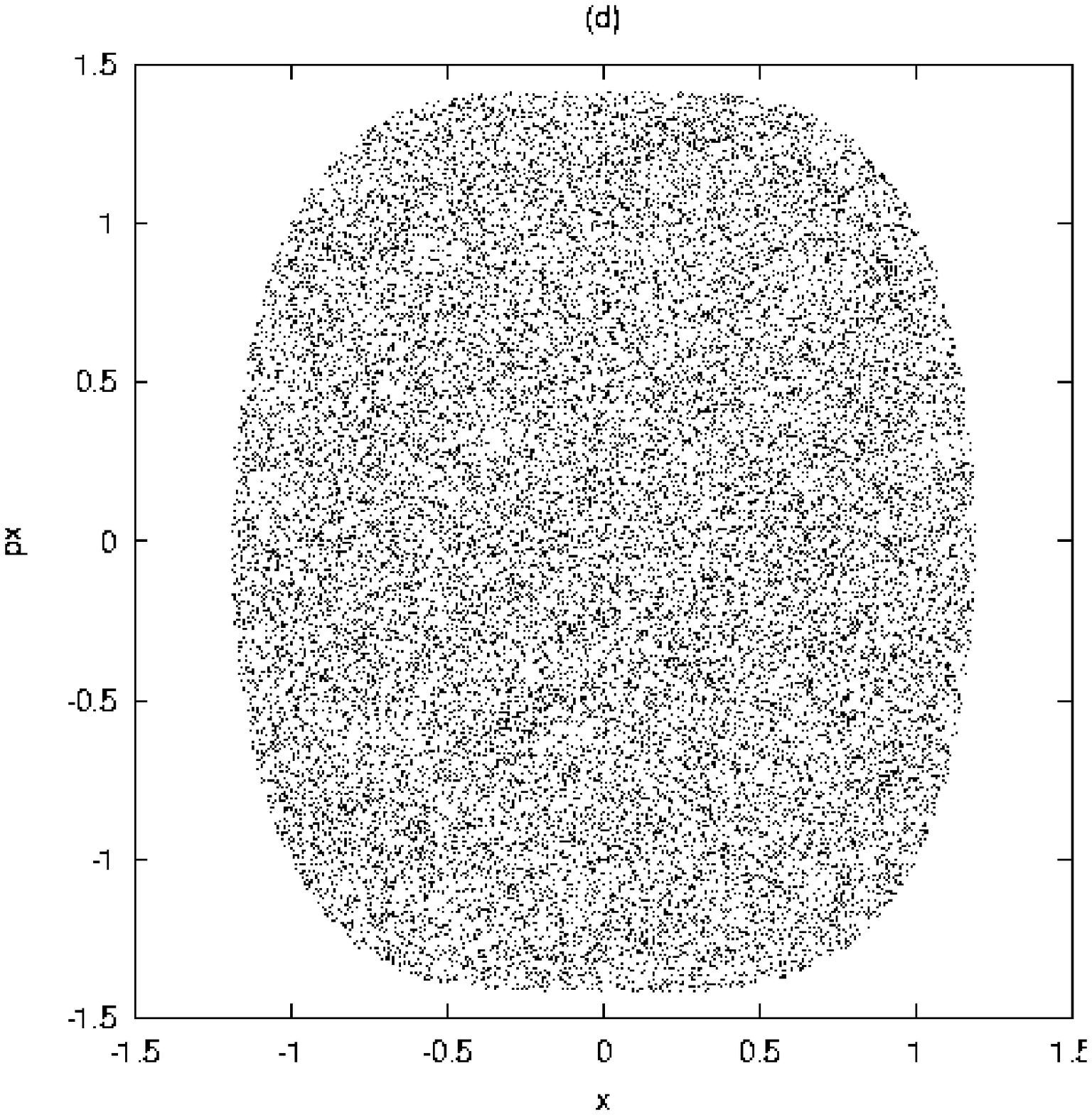}
\end{minipage}
\caption{Poincar\'{e} surface of section at $E=1$ and $y=0$ for
$k=0.0$ (a), $0.2$ (b), $0.4$ (c) and $0.6$ (d).}
\label{section}
\end{center}
\end{figure}

Fig.\ref{section} shows the Poincar\'{e} surfaces of section 
of this system at $k=0.0, 0.2, 0.4,$ and $0.6$. 
At $k=0.0$, the phase space is completely occupied by tori.
These tori are partially destroyed at $k=0.2$, and most
of them seem to vanish at $k=0.4$. However, there are still
many islands of stability. The two most significant islands are 
at $x=0$. These correspond to the stable linear orbits on 
$x=y$ and $x=-y$. At $k=0.6$, we can not recognize any structure.
In this case, more than 90\% of the orbits are unstable according
to \cite{meyer}.    

\subsection{Diagonalization of the Hamiltonian}

We diagonalize the Hamiltonian (\ref{hamiltonian}) in
the harmonic oscillator basis. The procedure we used is essentially
the same as that of Zimmermann {\it et al.} \cite{zimmermann}.
The bases belonging to $A_{1}$ 
can be written as 
\begin{eqnarray}
|n_{x},n_{y}\rangle_{A_{1}} = \frac{|n_{x},n_{y}\rangle + |n_{y},n_{x}\rangle}{\sqrt{2(1+\delta_{n_{x},n_{y}})}},
\end{eqnarray}
where $n_{x}$ and $n_{y}$ are even non-negative integer
and $n_{x}\ge n_{y}$.
Matrix elements of the Hamiltonian (\ref{hamiltonian}) 
can be calculated analytically, and we obtain eigenenergies
and eigenstates by the diagonalization of this matrix.
The truncated Hilbert space is spanned by the basis vectors with 
$0\le n_{x}+n_{y}\le 270$, whose dimension is 4692.

Among the 4692 eigenstates we obtain by the diagonalization, 
those having very large energies are not reliable because
of the truncation error. Since the range of the validity very much
depends on the oscillator frequency $\omega$ of the 
harmonic oscillator basis,
we should choose $\omega$ to optimize the diagonalization.
Because of the variational principle, eigenvalues obtained in
the restricted Hilbert space are always higher than the real values,
and the level density in that space is always smaller than
the real one. Therefore the minimization of ${\rm tr}H$ 
(the sum of the eigenvalues) can be
a criterion to choose the optimal $\omega$.

We put $\omega=7.0$, 
which is chosen roughly to optimize the diagonalization
for $k=0.6$, for all $k$. 
The comparison between the obtained level density
and the semiclassical one shows that the maximum reliable energy
is $E=800\sim 1000$.

\subsection{The second moment of the Husimi distribution and the number
of principal components}

\begin{figure}
\begin{center}
\includegraphics[height=0.5\textwidth]{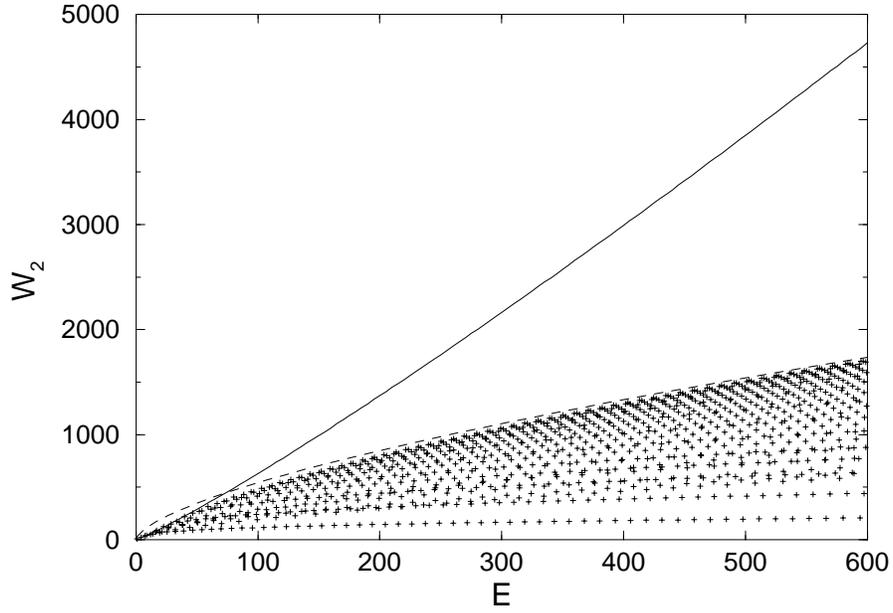}
\end{center}
\caption{Results of numerical calculation of $W_{2}$ 
at $k=0.0$.
The solid line shows the ergodic limit calculated based on 
the Berry-Voros hypothesis. The broken line shows the semiclassical upper
limit based on the torus quantization. See Appendix \ref{semi}.}
\label{im2k0}
\end{figure}

\begin{figure}
\begin{center}
\includegraphics[height=0.5\textwidth]{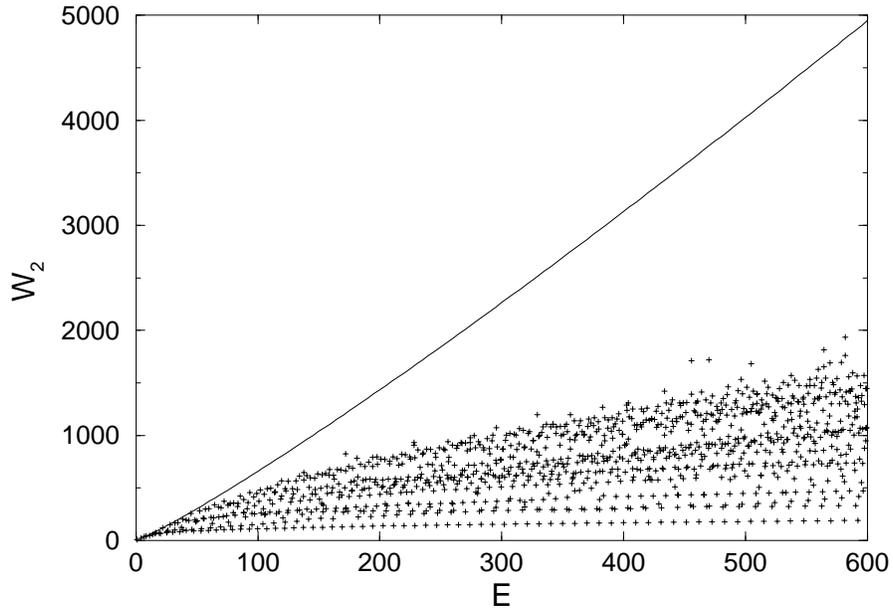}
\end{center}
\caption{$W_{2}$ at $k=0.2$.}
\label{im2k2}
\end{figure}

\begin{figure}
\begin{center}
\includegraphics[height=0.5\textwidth]{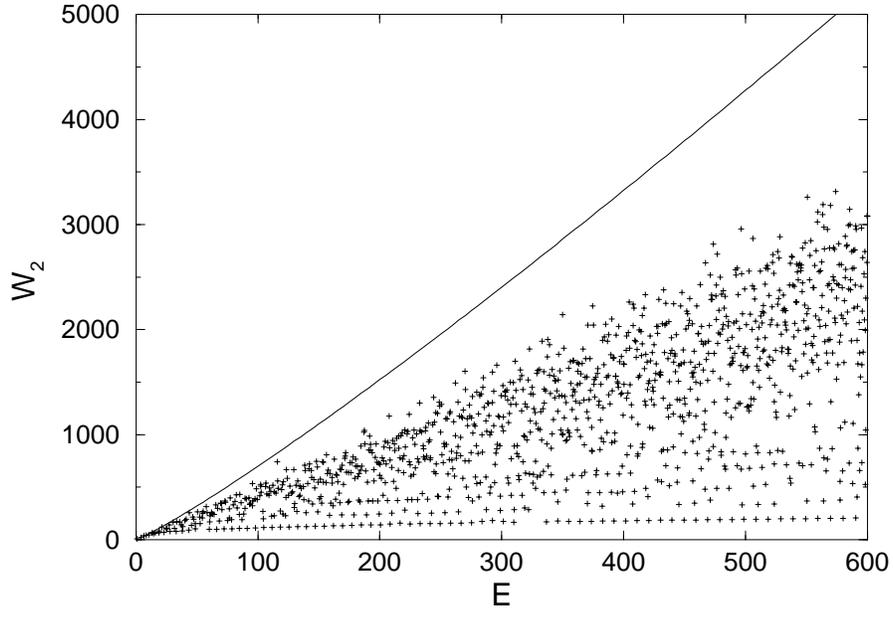}
\end{center}
\caption{$W_{2}$ at $k=0.4$.}
\label{im2k4}
\end{figure}

\begin{figure}
\begin{center}
\includegraphics[height=0.5\textwidth]{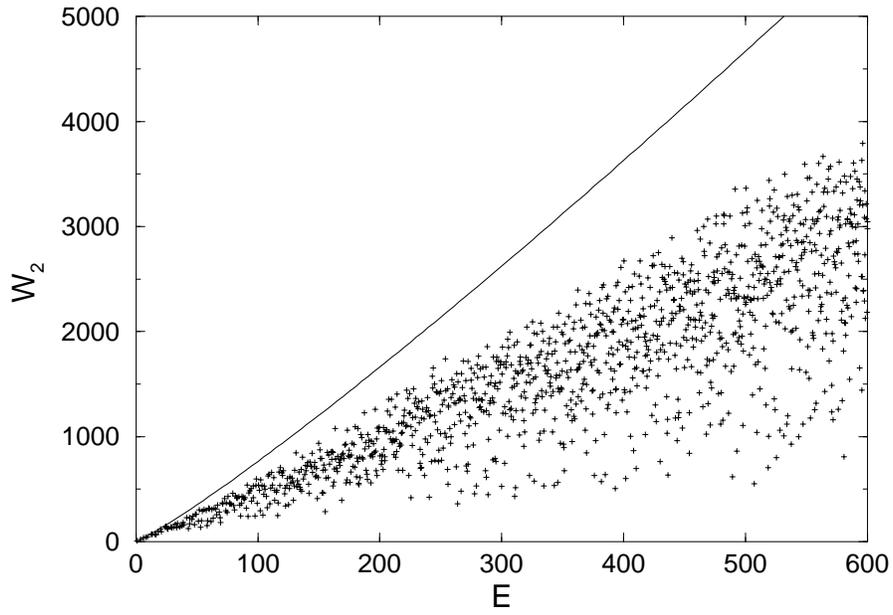}
\end{center}
\caption{$W_{2}$ at $k=0.6$.}
\label{im2k6}
\end{figure}

Figs. \ref{im2k0}, \ref{im2k2}, \ref{im2k4} and \ref{im2k6} show
the results of the numerical calculation of the inverse of
the second moment ($W_{2}$) of the Husimi distribution for $k=0.0, 0.2,
0.4$ and $0.6$. We plotted them for energy eigenstates with $E\le 600$.
In each figure, $1000 \sim 1500$ eigenstates are plotted.
It takes about 30 minutes on the NEC SX-5 in Osaka University
to calculate the second moments for
all (4692) eigenstates. 

When we calculate the Husimi distribution, $\lambda$ in (\ref{def})
is a free parameter. The simplest choice is to set $\lambda$
equal to the harmonic oscillator frequency of the basis 
used for the diagonalization. In this case, we can directly 
calculate the second moment using the formulas (\ref{general1})
and (\ref{general2}). 
The results in Figs. \ref{im2k0}, \ref{im2k2}, \ref{im2k4} 
and \ref{im2k6} are calculated using the optimized value 
$\lambda = 7.0$.

\begin{figure}
\begin{center}
\begin{minipage}{16cm}
\includegraphics[height=0.4\textwidth]{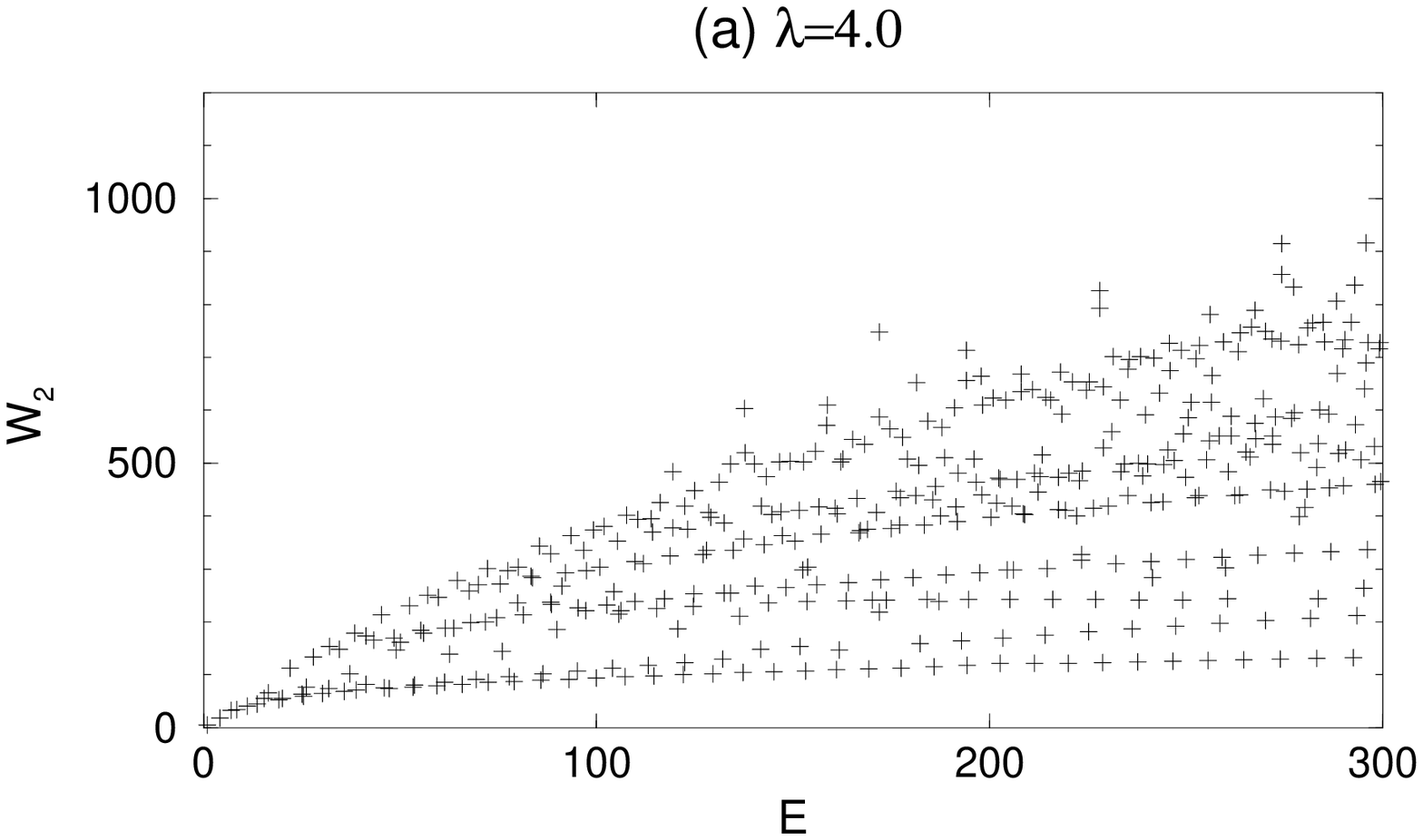}
\end{minipage}

\begin{minipage}{16cm}
\includegraphics[height=0.4\textwidth]{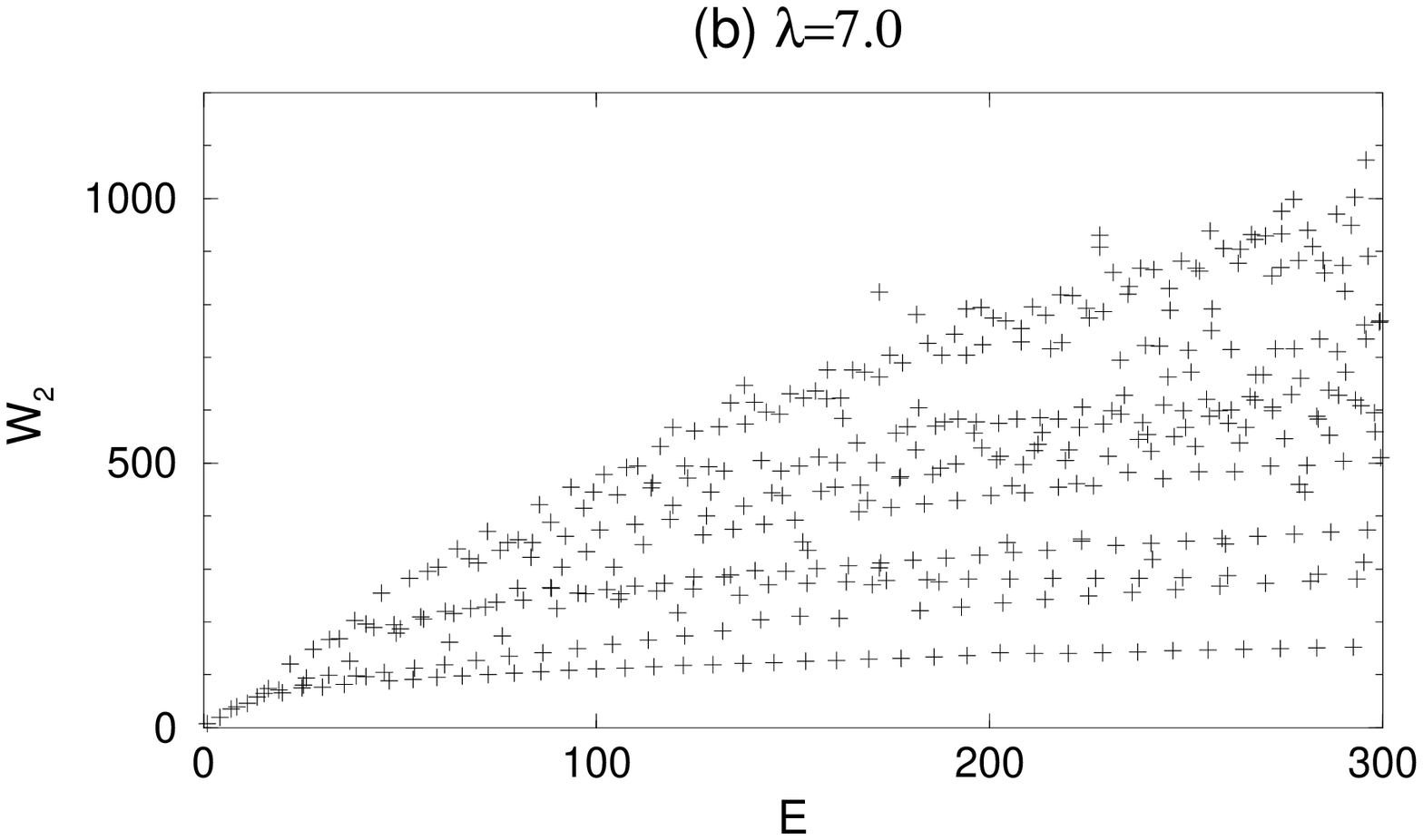}
\end{minipage}

\begin{minipage}{16cm}
\includegraphics[height=0.4\textwidth]{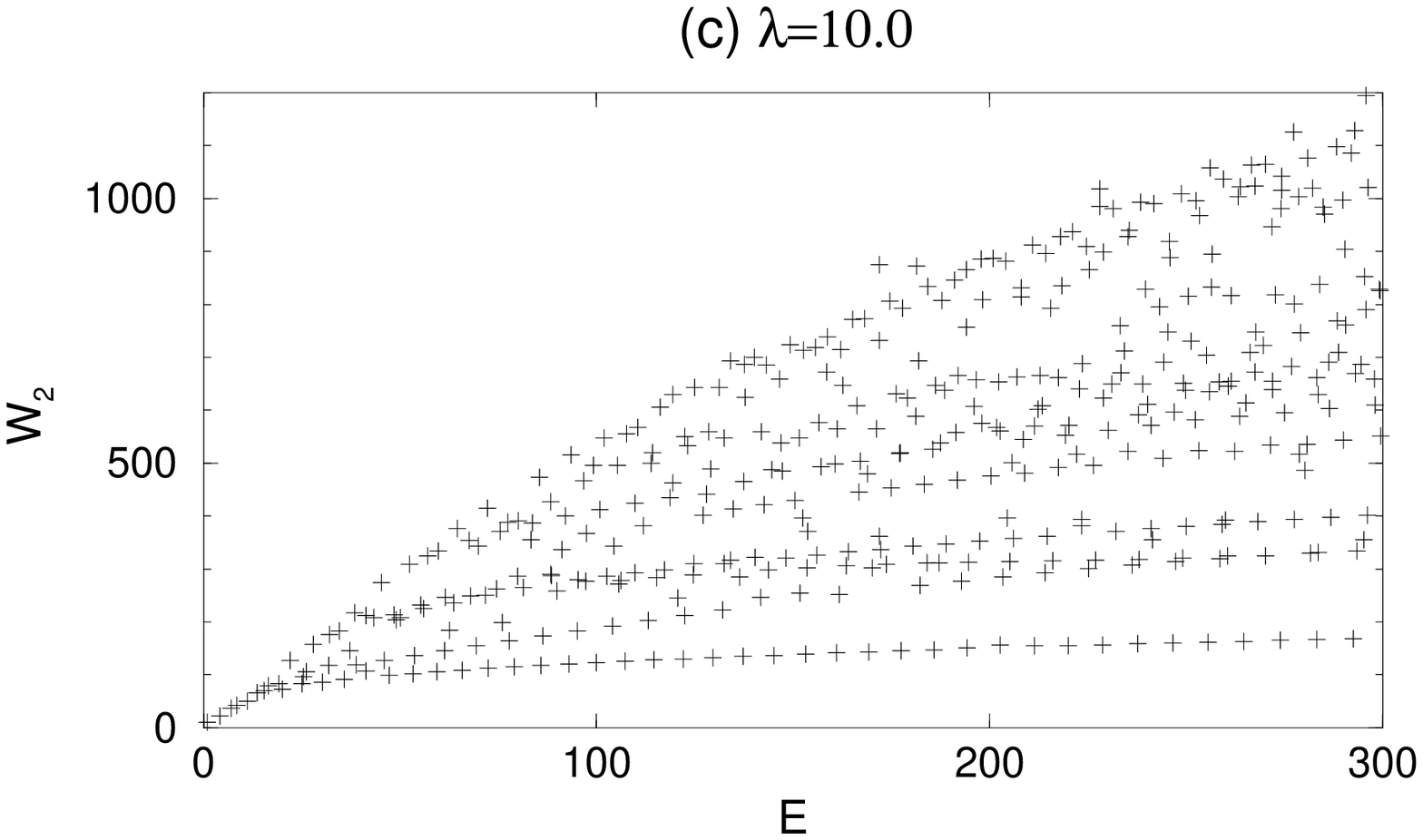}
\end{minipage}
\end{center}
\caption{$W_{2}$ at $k=0.2$ for $\lambda =4.0$ (a), 
$7.0$ (b) and $10.0$ (c). 
The values of $W_{2}$ depend on $\lambda$, but the qualitative
features remain unchanged.}
\label{lambda}
\end{figure}

We also calculated $W_{2}$ for different values of $\lambda$.
Fig. \ref{lambda} shows the results for $\lambda = 4.0, 7.0$ and $10.0$
at $k=0.2$. Qualitative features of the figures 
seem unchanged, at least when  
$\lambda$ is not so far from
the optimized value. When  
$\lambda$ is far from the optimized value, it is hard to obtain
reliable results because the number of basis vectors we need 
to represent the eigenstates is huge.

At $k=0.0$, the system is separable into two
one-dimensional systems. 
Therefore eigenstates of the original system
can be specified by quantum numbers 
$m_{x}$ and $m_{y}$, which label eigenstates
of the one-dimensional systems.

\begin{figure}
\begin{center}
\includegraphics[width=1.0\textwidth]{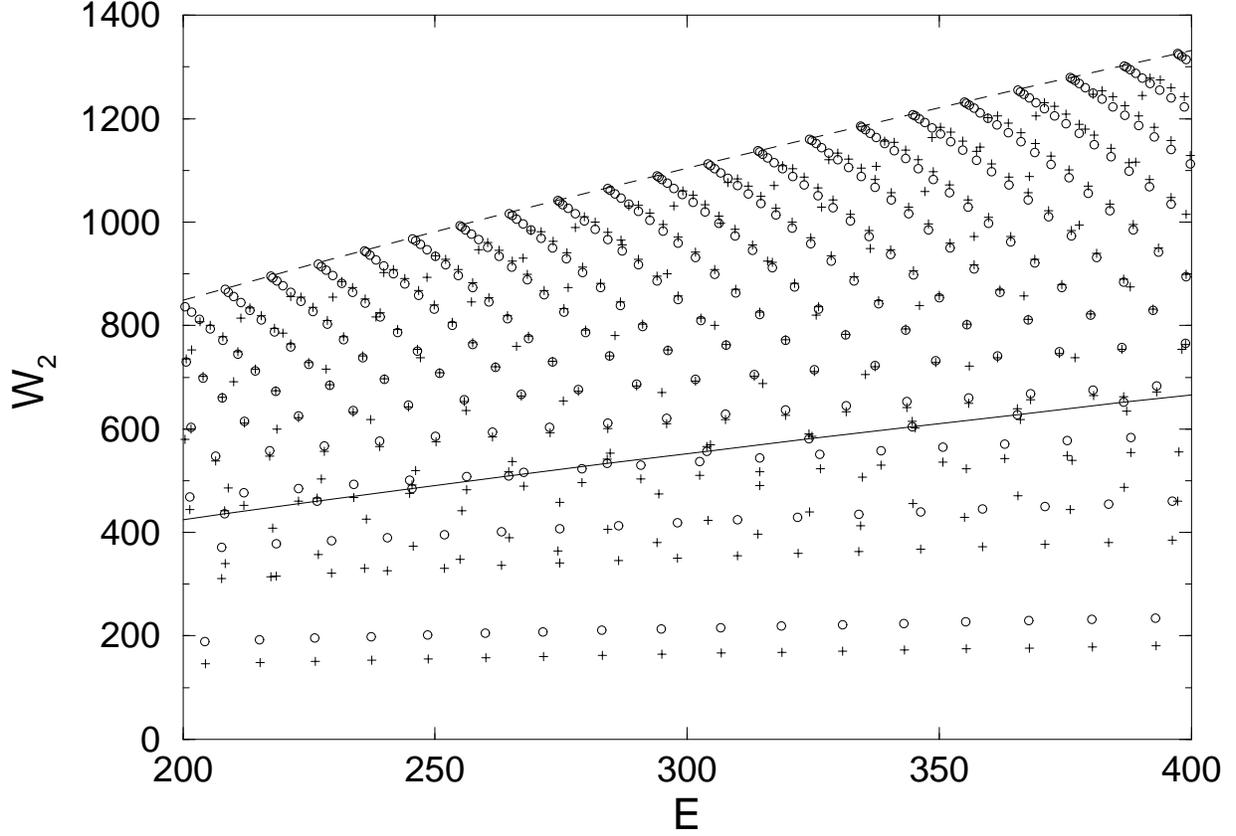}
\end{center}
\caption{Comparison between quantum results and semiclassical 
results based on torus quantization at $k=0.0$. Pluses are the
quantum results and circles are the semiclassical results. 
The broken
line is the upper limit (\ref{w2bar}), 
and the solid line is exactly one half thereof. Eigenstates with $m_{x}=m_{y}$
are located on the solid line. In the upper half of this figure, pluses
and circles have a good correspondence. However, very near the 
broken line, the correspondence is lost because the assumption 
(\ref{separate}) is not good when $m_{x}\sim m_{y}$. In the lower
part of this figure, there are eigenstates with small $m_{y}$, and 
semiclassical values are a little higher than the exact values.}
\label{BS}
\end{figure}

We can assume $m_{x}\ge m_{y}$ without loss of generality
in the class $A_{1}$.
Fig.\ref{BS} shows the results of a semiclassical calculation based on
the torus quantization. (For details of the calculation, 
see Appendix \ref{torus}.) Semiclassical results
and full
quantum results have a good correspondence, 
except for the cases
where $m_{x}\sim m_{y}$ and $m_{y}\sim 0$. When 
$m_{x}\sim m_{y}$, the assumption (\ref{separate}) does not
seem to apply. When $m_{y}\sim 0$, the quantum number is too
small to use the semiclassical approximation. 

\begin{figure}
\begin{center}
\includegraphics[height=1.0\textwidth]{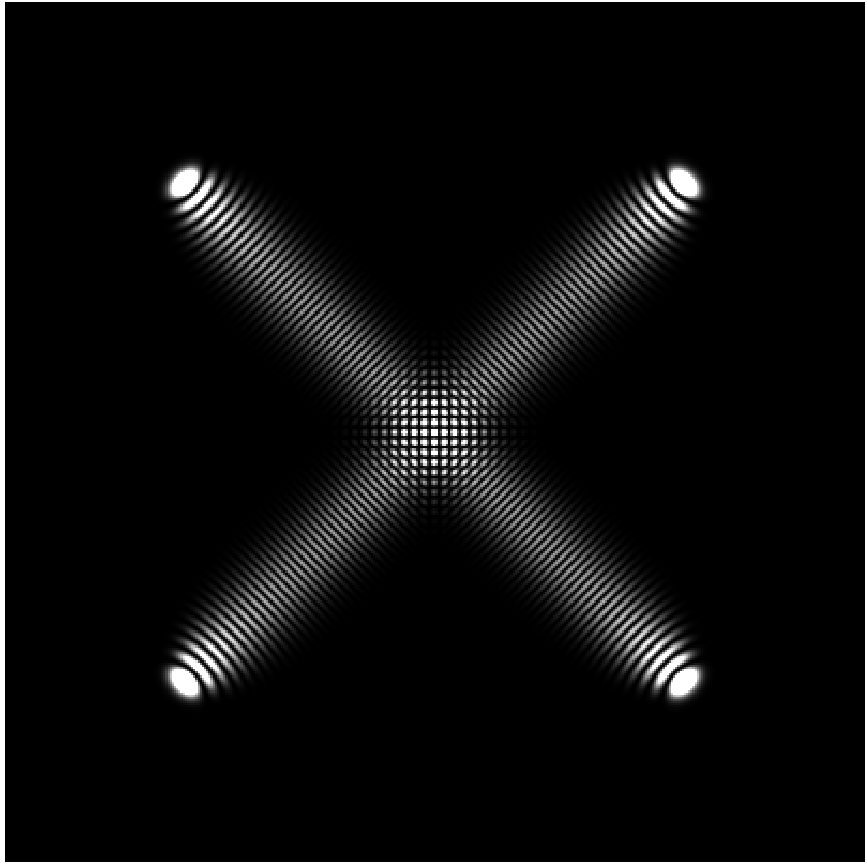}
\end{center}
\caption{The density plot of the square of the 441st 
eigenfunction at $k$=0.2.   
$E$=320.8, $W_{2}$=157.2, NPC=102.8. 
The length of the sides of this
figure is 15. 
This state is completely localized around 
the diagonal orbits, and there is
no node in the transverse direction.}
\label{w441k2}
\end{figure}

In Fig.\ref{im2k2}, the regular structure of Fig.\ref{im2k0}
is partially destroyed, but there are still many regular series
of eigenstates. The most significant series, which is in the lowest 
part of the Fig.\ref{im2k2}, corresponds to the stable 
diagonal periodic orbits
at $x=y$ and $x=-y$. Fig.\ref{w441k2} shows an eigenstate
in the regular series. 
There are also some series of 
regular eigenstates which are excited in the transverse direction
to the periodic orbits. 

The lowest series in Fig.\ref{im2k0} corresponds to the
periodic orbits at $x=0$ and $y=0$, which are stable 
at $k=0.0$. However, these orbits are unstable at
$k>0$, and this series seems to disappear as $k$
increases.

The eigenstates corresponding to the
torus with $E_{x}=E_{y}$ 
are in the middle of Fig.\ref{im2k0}. (See also Fig.\ref{BS}.) 
As $k$ increases, this torus is quickly destroyed and two stable
diagonal orbits are left. Islands of stability around 
these two orbits become the most significant structure in this system, 
and the eigenstates localized around these
orbits form the lowest series in Fig.\ref{im2k2}. 

At $k=0.4$, as seen from Fig.\ref{im2k4}, 
there are still regular series that correspond
to the diagonal orbits and the first excited states in the transverse
direction. However, most structures seem to have been destroyed and many
eigenstates are near the ergodic limit, as based
on the Berry-Voros hypothesis. (See Appendix \ref{ergodic}.)  

At $k=0.6$, no clear structure can be seen in Fig.\ref{im2k6},
and the values of $W_{2}$ go up as a whole.
However, the ergodic limit based on the Berry-Voros hypothesis
is still not reached, and some eigenstates have much smaller
values than the limit.

There may be several reasons for this.
One reason we can think of is that the Wigner function in the
Berry-Voros hypothesis does not satisfy the pure state condition of the
density matrix
$\hat{\rho}^{2}=\hat{\rho}$.
Therefore we might obtain better estimates of $W_{2}$ by taking
into account the pure state condition by successive iterations.
(See chapter 8 in \cite{almeida}.) 
Another reason, which seems more important, is that 
there are weak localization phenomena like {\it scars}
\cite{heller}. We found many scarred eigenstates with
small $W_{2}$ at $k=0.6$.

At low energies ($E<50$), $W_{2}$ of some eigenstates
reach the ergodic limit irrespective of the value of $k$,
but the figures of these wave functions does not seem chaotic. 
The reason for this is probably that the semiclassical approximation
we used here is not good in this region. 
Our approximation is based on the idea that the local volume
occupied by the Husimi distribution is
the volume element of the invariant manifold 
(invariant torus or equi-energy surface)
multiplied by the thickness factor of order $\hbar^{l/2}$, where $l$ is 
the codimension of the manifold. (See Appendix \ref{semi}.)
However, this approximation 
fails when the scale of the invariant manifold is comparable
to the Planck constant.
For example, in the low energy part of Fig. \ref{im2k0}, 
the ergodic limit line is under the limit based 
on the torus quantization.
There seems no clear distinction between chaotic and regular 
eigenstates at very low energies.

\begin{figure}
\begin{center}
\begin{minipage}{8cm}
\includegraphics[height=1.0\textwidth]{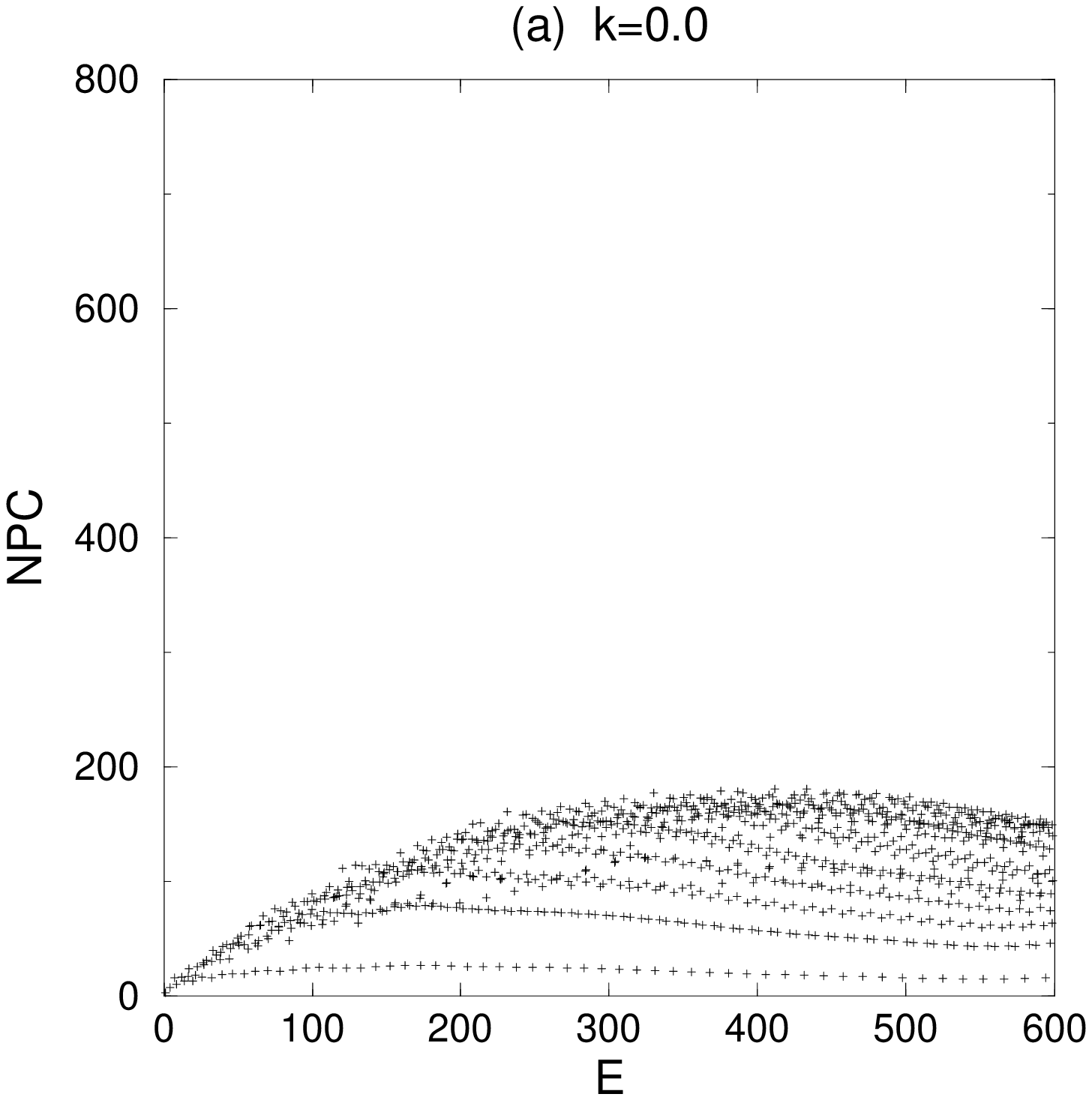}
\end{minipage}
\begin{minipage}{8cm}
\includegraphics[height=1.0\textwidth]{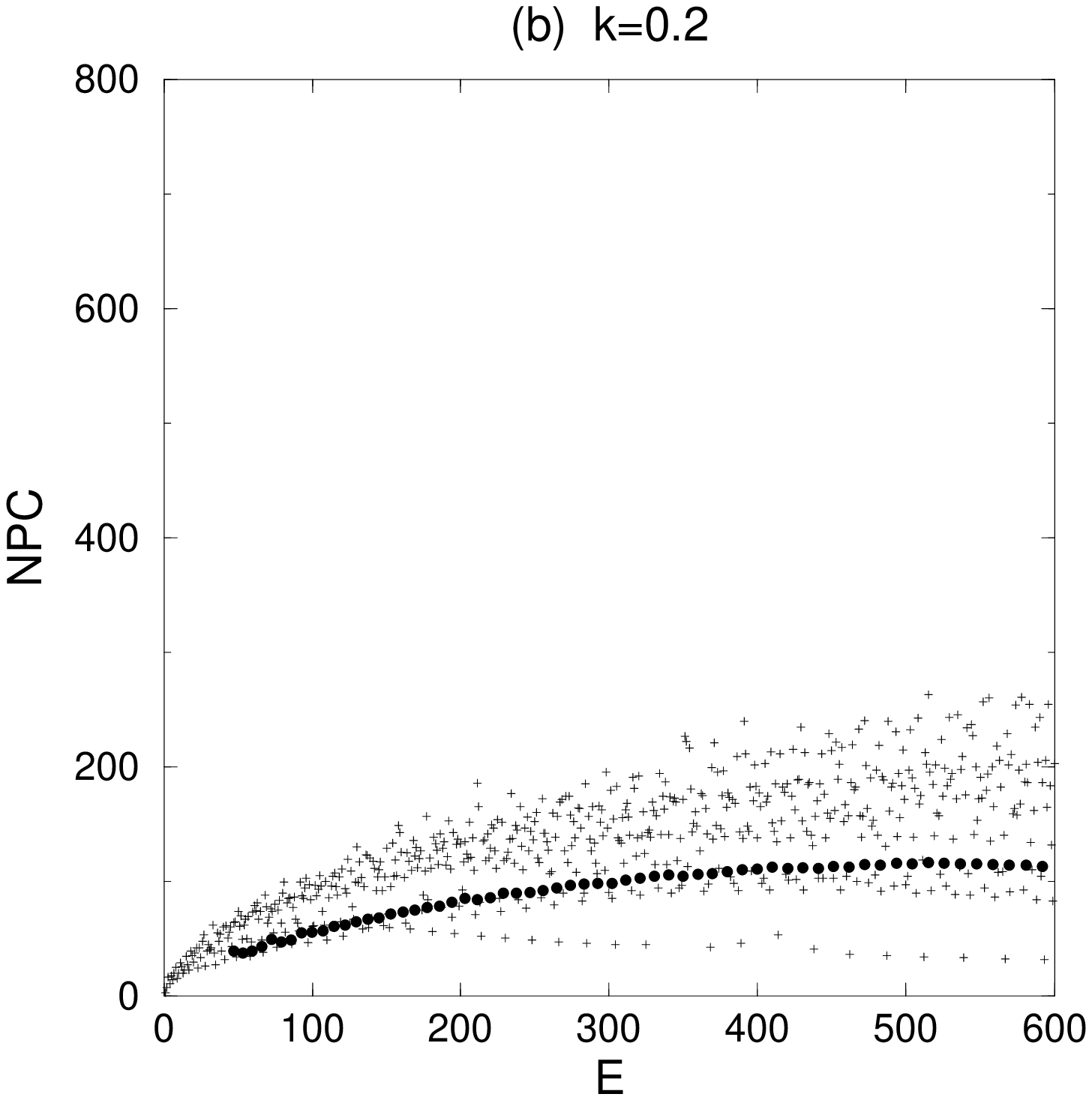}
\end{minipage}

\begin{minipage}{8cm}
\includegraphics[height=1.0\textwidth]{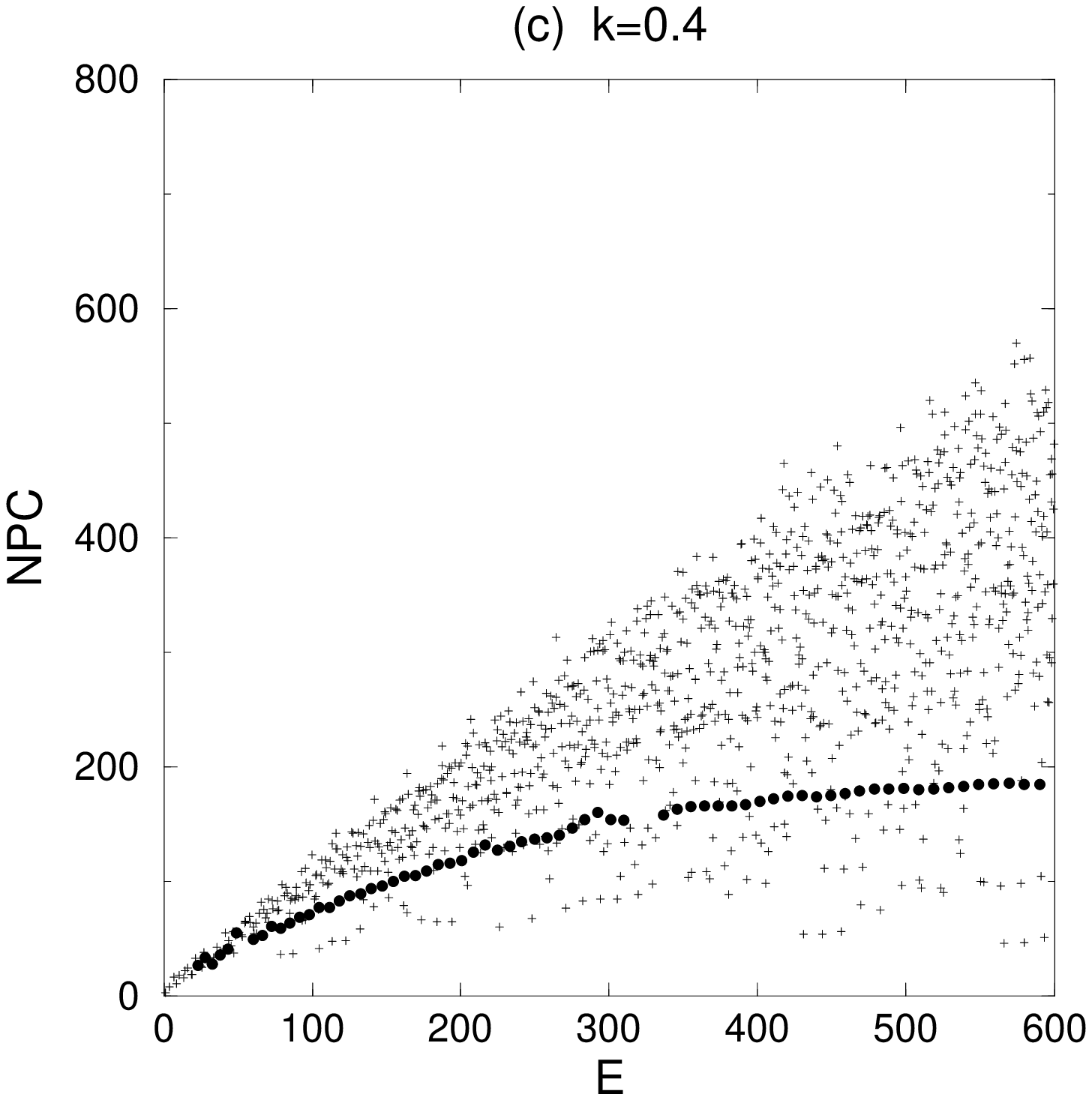}
\end{minipage}
\begin{minipage}{8cm}
\includegraphics[height=1.0\textwidth]{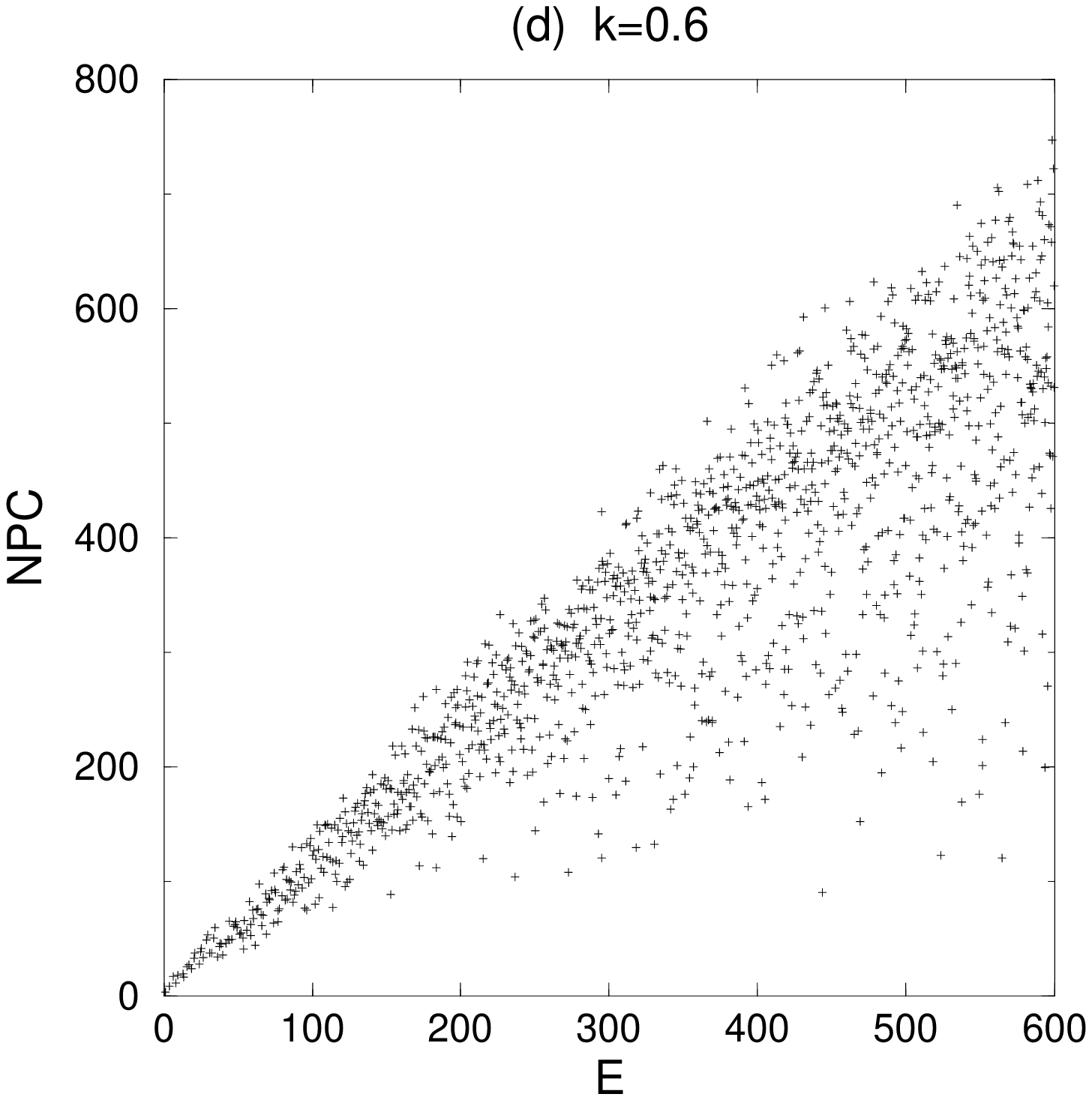}
\end{minipage}
\end{center}
\caption{NPC at $k=0.0$ (a), $0.2$ (b), $0.4$ (c) and
$0.6$ (d). Dots in (b) and (c) show the regular series
corresponding to the diagonal orbits.}
\label{NPC}
\end{figure}

\begin{figure}
\begin{center}
\includegraphics[height=1.0\textwidth]{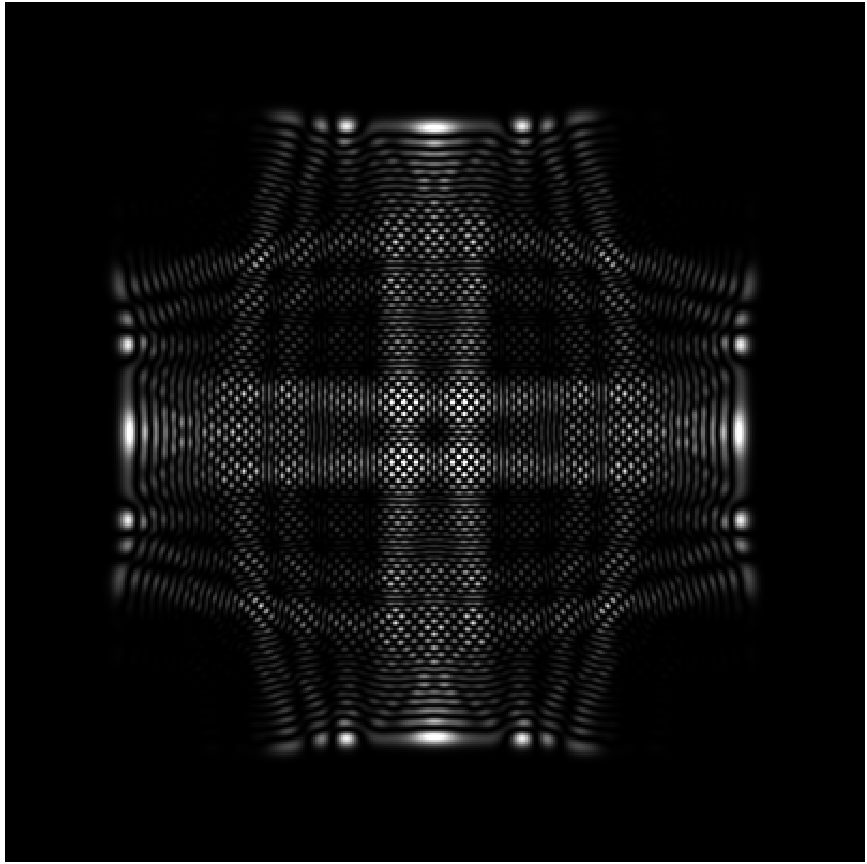}
\end{center}
\caption{754th eigenstate at $k$=0.4. 
$E$=443.6, $W_{2}$=545.1, NPC=54.1. 
NPC of this state
is very small, though $W_{2}$ is not so small and its structure
is not so clear. We can see scars of linear orbits at
$x=0$ and $y=0$.}
\label{w754k4}
\end{figure}

In Fig. \ref{NPC}, 
we plotted the NPC at $k=0.0, 0.2, 0.4$ and $0.6$. 
We can see regular structure in Fig.\ref{NPC} (a). The structure
is gradually broken as $k$ increases, and the points are lifted
as a whole. In that sense, the behavior of NPC is similar to
that of $W_{2}$. However, by comparing NPC and $W_{2}$ of each
eigenstate, we can see that the two quantities are not so similar. 
For instance, at $k=0.4$, we can see regular series
of eigenstates related to the diagonal orbits in both Fig. \ref{im2k4}
and \ref{NPC} (c). However, the values of the NPC are
not so small, and there
are many states lower than the series which seem to have no clear
structure. Therefore, in general, $W_{2}$ seems more reliable than 
the NPC as
a measure of complexity of quantum states. Some eigenstates 
that have small NPC are related to the scars of the 
unstable linear orbits at $x=0$ and $y=0$. (See Fig. \ref{w754k4}.)
Importance of these eigenstates
in the response function
has been reported in \cite{aiba}. 

\section{Summary}\label{summary}

In this paper we proposed the second moment of the Husimi
distribution as a definition of complexity of quantum states.
Its inverse (we represent it by $W_{2}$) shows the effective
volume occupied by the Husimi distribution function, 
and serves as a measure of delocalization in phase space.
We calculated it for a quartic oscillator model, and showed that
it has a good correspondence with chaoticity of the classical system.
We can calculate it directly from expansion coefficients without
numerical integration.
Therefore the calculation is not so time-consuming, and
there is no numerical error except for that contained 
in the quantum state itself. 

In the integrable case ($k=0.0$), the values of $W_{2}$ have a
regular structure. As $k$ increases, the structure is gradually
destroyed and the values go up near the ergodic limit line as
a whole. At $k=0.2$ and $0.4$, there are many regular series
related to stable islands in classical phase space. Even at
$k=0.6$, the values of $W_{2}$ of some eigenstates are much lower than
the ergodic limit based on the Berry-Voros hypothesis, 
and they seem to be related to scars of unstable
periodic orbits.

To generalize the idea of this paper to many-body systems
using generalized coherent states is interesting.  
It will be reported in the forthcoming paper \cite{sugita}.

\section*{Acknowledgments}
The authors thank Prof. V. Dmitra\v{s}inovi\'{c} and 
Dr. E. -M. Ilgenfritz for reading
the manuscript and making a number of helpful suggestions.
They thank also the members of the Nuclear Theory Group
at Kyoto University for valuable discussions.
The numerical calculations have been done on NEC SX-5 at Research
Center for Nuclear Physics (RCNP), Osaka University. 
A.S. is supported by the center-of-excellence (COE) program at
RCNP.

\appendix
\section{Semiclassical calculations}\label{semi}
In this appendix we give, for comparison, semiclassical
estimates for the second moment $M_{2}$.

\subsection{Integrable case ($k=0$)}\label{torus}

In this case, the model Hamiltonian is separable:
\begin{eqnarray}
H(p_{x},p_{y},x,y) = h(p_{x},x) + h(p_{y},y), 
\label{2dim}
\end{eqnarray}
\begin{eqnarray}
h(p,x) = \frac{1}{2}\left(p^{2} + x^{4}\right).
\label{h1dim}
\end{eqnarray}
Therefore, we first derive a semiclassical formula for
the Husimi function
of eigenstates of (\ref{h1dim}).  
After that, it is easy to construct the semiclassical eigenstates of 
(\ref{2dim}) taking into account
discrete symmetries.

\subsubsection{One-dimensional problem}
The action variable of the one-dimensional system is 
\begin{eqnarray}
I & = & \frac{1}{2\pi}\oint p dx, \\
& = & 
\frac{4}{3\pi}(2E)^{3/4}C,
\end{eqnarray}
\begin{eqnarray}
C = \int_{0}^{1}\frac{dx}{\sqrt{1-x^{4}}}
= \frac{1}{4\sqrt{2\pi}}\Gamma\left(\frac{1}{4}\right)^{2}.
\end{eqnarray}
The semiclassical eigenvalues are determined by the Bohr-Sommerfeld
quantization condition
\begin{eqnarray}
I(E_{m}) = (2m+1)\pi\hbar,
\end{eqnarray}
i.e.
\begin{eqnarray}
E_{m} = \frac{1}{2}\left(\frac{3\pi}{4C}\right)^{4/3}
\left(m+\frac{1}{2}\right)^{4/3}.
\label{bohr-sommerfeld}
\end{eqnarray}
The semiclassical Wigner function of the eigenstate is
\begin{eqnarray}
\rho_{W}(p,q) = 
\frac{1}{2\pi} \delta\left(I\left(E(p,q)\right)-I(E_{m})\right),
\end{eqnarray}
and the Husimi function is obtained by Gaussian smearing thereof:
\begin{eqnarray}
& &\rho_{H,\lambda}(p,q) \\
& = & \frac{1}{\pi\hbar}
\int dp'dq' \exp\left[\frac{-1}{\hbar}
\left\{\lambda (q-q')^{2}+\frac{(p-p')^{2}}{\lambda}\right\}\right]
\rho_{W}(p',q'). 
\end{eqnarray}
By changing coordinates $q\rightarrow q/\sqrt{\lambda}, q'\rightarrow
q'\sqrt{\lambda}, p\rightarrow\sqrt{\lambda}p,
p'\rightarrow\sqrt{\lambda}p'$, this becomes equivalent 
to using the modified Hamiltonian
\begin{eqnarray}
H = \frac{\lambda}{2}p^{2} + \frac{1}{2\lambda^{2}}q^{4},
\end{eqnarray}
and putting parameter $\lambda = 1$ in the Gaussian.

In this case, the Husimi function near the equi-energy surface
can be written approximately
\cite{takahashi}
\begin{eqnarray}
\rho_{H}(s,\xi) = \frac{2\sqrt{\pi\hbar}}{T|grad H(s)|}
\exp\left[-\frac{(\xi-\xi_{0})^{2}}{\hbar}\right].
\end{eqnarray}
Here, $s$ is a coordinate which parameterizes
the equi-energy surface. $\xi$ is the other
coordinate, and $\xi=\xi_{0}$ corresponds to a point on the surface.
$s$ and $\xi$ are assumed to be orthonormal. 
$T$ is the period, whose explicit form is
\begin{eqnarray}
T = 4C(2E)^{-1/4}.
\end{eqnarray} 

To calculate $M_{2}(\rho_{H})$, we introduce new coordinates
$(E,\theta)$ as
\begin{eqnarray}
p & = & \sqrt{\frac{2E}{\lambda}}\sin\theta, \\
q & = & \left(2E\right)^{1/4}\sqrt{\lambda}\cos^{1/2}\theta.
\end{eqnarray}
Then
\begin{eqnarray}
M_{2}(\rho_{H}) & = & 
\int \frac{d\xi ds}{2\pi\hbar} \rho_{H,x}(\xi,s)^{2}, \\
& = &
\frac{\sqrt{2\pi\hbar}}{T^{2}}\int d\theta 
\left|\frac{ds}{d\theta}\right| \frac{1}{|grad E(\theta)|^{2}},\\
& = &
\frac{1}{32C^{2}}\sqrt{\frac{2\pi\hbar}{E}}
\int_{0}^{\pi/2}\frac{d\theta}{\sqrt{\frac{1}{\lambda}\cos^{4}\theta 
+ \frac{\lambda}{4\sqrt{2E}}\sin^{2}\theta\cos\theta}}.
\label{m2torus}
\end{eqnarray}

\subsubsection{Two-dimensional solutions}
Since the model (\ref{hamiltonian}) has some discrete symmetries,
we have to take them into account when we construct semiclassical eigenstates.
For example, in the class $A_{1}$, an eigenstate is written as
\begin{eqnarray}
|m_{x},m_{y}\rangle_{A_{1}} = 
\frac{|m_{x},m_{y}\rangle +|m_{y},m_{x}\rangle}
{\sqrt{2(1+\delta_{m_{x},m_{y}})}},
\end{eqnarray}
where $|m_{x},m_{y}\rangle $ is the product of one-dimensional
eigenstate of (\ref{h1dim}), and both $m_{x}$ and $m_{y}$ are
even.

If $m_{x}\ne m_{y}$, the Husimi function of the eigenstate is 
\begin{eqnarray}
|\langle z |m_{x},m_{y}\rangle_{A_{1}}|^{2} = 
\frac{1}{2}\left(
|\langle z |m_{x},m_{y}\rangle |^{2} +
|\langle z |m_{y},m_{x}\rangle |^{2} + 
2{\rm Re} \langle m_{x},m_{y}|z\rangle\langle z|m_{y},m_{x}\rangle
\right).
\end{eqnarray}
In the semiclassical approximation, $|m_{x},m_{y}\rangle $ and
$|m_{y},m_{x}\rangle$ correspond to different tori.
Therefore we assume 
\begin{eqnarray}
\langle m_{x},m_{y}|z\rangle\langle z|m_{y},m_{x}\rangle \sim 0 .
\label{separate}
\end{eqnarray}
Under this assumption, 
\begin{eqnarray}
\rho_{H}(|m_{x},m_{y}\rangle_{A_{1}}) =
\left\{
\begin{array}{cc}
\frac{1}{2}\left(\rho_{H}(|m_{x},m_{y}\rangle) +
\rho_{H}(|m_{y},m_{x}\rangle)\right) 
& (m_{x}\ne m_{y}) \\ 
\rho_{H}(|m_{x},m_{y}\rangle )
& (m_{x} = m_{y})
\end{array}
\right. .
\end{eqnarray}
The second moments are 
\begin{eqnarray}
M_{2} = \left\{
\begin{array}{cc}
\frac{1}{2}M_{2}(E_{m_{x}})M_{2}(E_{m_{y}}) 
& (m_{x}\ne m_{y}) \\ 
M_{2}(E_{m_{x}})M_{2}(E_{m_{y}})
& (m_{x} = m_{y})
\end{array}
\right. .
\end{eqnarray}
Here, $M_{2}(E)$ denotes the second moment of the one-dimensional
problem at energy $E$, whose explicit form is given in (\ref{m2torus}). 
$E_{n}$ is the $n$-th eigenenergy of the one-dimensional
problem, as given in (\ref{bohr-sommerfeld}).

If we change 
$E_{x}$ and $E_{y}$ fixing the total energy $E=E_{x}+E_{y}$,
the product $M_{2}(E_{x})M_{2}(E_{y})$
takes its minimum value at $E_{x}=E_{y}$ 
because $M_{2}(E)$ behaves like $E^{\alpha}$ with 
$-1/2 \le \alpha \le -1/4$.
Therefore the upper limit of $W_{2}$ with fixed $E$ 
(we denote this as $\bar{W}_{2}(E)$) is 
\begin{eqnarray}
\bar{W}_{2}(E) = \frac{2}{M_{2}\left(\frac{E}{2}\right)^{2}}.
\label{w2bar}
\end{eqnarray} 
$W_{2}$ of the eigenstates with $m_{x}\sim m_{y}$ are close
to $\bar{W}_{2}$, but if $m_{x}=m_{y}$, $W_{2}$ of the state is
one half of $\bar{W}_{2}$.

\subsection{Ergodic limit}\label{ergodic}

In this subsection, we calculate the Husimi distribution corresponding
to the Berry-Voros hypothesis \cite{berry,voros} and the second moment
thereof.

The Berry-Voros conjecture can be stated as saying 
that the Wigner function for the
stationary state of an ergodic system is approximately
\begin{eqnarray}
\rho_{W}(\bp,\bq) = N_{W}\delta\{E-H(\bp,\bq)\},
\label{berry-voros}
\end{eqnarray}
where the normalization is, in $k$ dimensions,
\begin{eqnarray}
N_{W} =
\left[\int \frac{d\bp d\bq}{(2\pi\hbar)^{k}}
\delta\{E-H(\bp,\bq)\}\right]^{-1}.
\end{eqnarray} 
The Husimi function $\rho_{H,\lambda}$ is obtained by smearing 
(\ref{berry-voros}) by the Gaussian 
$e^{-\lambda \bq^{2} - \bp^{2}/\lambda}$.
From the discussion in the previous section, we can see that
this is equivalent to using the modified Hamiltonian
\begin{eqnarray}
H = \frac{\lambda}{2}(p_{x}^{2}+p_{y}^{2}) + 
\frac{1}{2\lambda^{2}}(x^{4}+y^{4}) - \frac{k}{\lambda^{2}}x^{2}y^{2},
\end{eqnarray} 
and putting $\lambda=1$ in the Gaussian.

We use coordinates $(E,\theta_{1},\theta_{2},\theta_{3})$, which
are related to original coordinates by the following equations:
\begin{eqnarray}
p_{x} & = & \sqrt{\frac{2E}{\lambda}}\cos\theta_{1}\cos\theta_{2}, \\
p_{y} & = & \sqrt{\frac{2E}{\lambda}}\cos\theta_{1}\sin\theta_{2}, \\
x & = &
\frac{E^{1/4}\sqrt{\lambda}}{f(\theta_{3})}
\sin^{1/2}\theta_{1}\cos\theta_{3},\\
y & = &
\frac{E^{1/4}\sqrt{\lambda}}{f(\theta_{3})}
\sin^{1/2}\theta_{1}\sin\theta_{3},
\end{eqnarray}
where
\begin{eqnarray}
f(\theta) & = & 
\left\{\frac{1}{2}(\cos^{4}\theta + \sin^{4}\theta) 
- k\cos^{2}\theta\sin^{2}\theta\right\}^{1/4}.
\end{eqnarray}

The normalization constant of
the Wigner function $N_{W}$ is determined as
\begin{eqnarray}
N_{W}^{-1} & = & \int \frac{d\bp d\bq}{(2\pi\hbar)^{2}}
\delta(E-H(\bp,\bq)), \\
& = &
\frac{4\pi\sqrt{E}}{(2\pi\hbar)^{2}}
\int_{0}^{\pi/2}\frac{d\theta}{f(\theta)^{2}}, \\
& = &
\frac{\sqrt{2E}}{\pi\hbar^{2}}
K\left(\sqrt{\frac{1+k}{2}}\right).
\end{eqnarray}
Here, $K$ is the complete elliptic integral. The Husimi function 
in this case is 
\begin{eqnarray}
\rho_{H} (\bp,\bq) & = & 
\frac{N_{H}}{|grad H(\bsigma)|}
\exp\left[\frac{-1}{\hbar}(\xi-\xi_{0})^{2}\right],
\end{eqnarray}
where $N_{H} = N_{W}/\sqrt{\pi\hbar}$ is the normalization
constant. $\bsigma$ is the coordinate which parameterizes
the equi-energy
surface, and $\bsigma$ and $\xi$ are orthonormal coordinates. 
The second moment of $\rho_{H}$ is
\begin{eqnarray}
M_{2}(\rho_{H}) & = &
\int\frac{d\xi d\bsigma}{(2\pi\hbar)^{2}}\rho_{H}(\xi,\bsigma)^{2}, \\
& = &
\frac{N_{H}^{2}}{(2\pi\hbar)^{2}}\sqrt{\frac{\pi\hbar}{2}}
\int \left|\frac{\partial \bsigma}{\partial\btheta}\right|
\frac{d\btheta}{|grad H(\btheta)|^{2}}. \\
\end{eqnarray}
Here, 
\begin{eqnarray}
|grad H(\btheta)|^{2} 
& = &
2E\cos^{2}\theta_{1} + \nonumber \\
&&
\frac{4E^{3/2}}{f(\theta_{3})^{6}}\sin^{3}\theta_{1} 
\left\{\cos^{6}\theta_{3}+\sin^{6}\theta_{3}
+ K(K-2)\cos^{2}\theta_{3}\sin^{2}\theta\right\},
\end{eqnarray}
and 
$\left|\frac{\partial \bsigma}{\partial\btheta}\right|d\btheta$ 
is the volume of a small three-dimensional region formed by
$(d\theta_{1},d\theta_{2},d\theta_{3})$. After lengthy, but
straightforward calculation we obtain
\begin{eqnarray}
\left|\frac{\partial \bsigma}{\partial\btheta}\right| =
\frac{E^{2}}{2f(\theta_{3})^{2}}\cos^{4}\theta_{1} +
\frac{4E^{5}}{f(\theta_{3})^{4}}\cos^{2}\theta_{1}\sin^{2}\theta_{1}
\left\{f'(\theta_{3})^{2} + f(\theta_{3})^{2}\right\}.
\end{eqnarray}
Finally we obtain the following formula
\begin{eqnarray}
M_{2}(\rho_{H}) & = &  \frac{N_{H}^{2}\pi\sqrt{E}}{2(2\pi\hbar)^{3/2}}
\int_{0}^{2\pi}d\theta_{3}
\int_{0}^{\pi/2}\frac{\cos\theta_{1}d\theta_{1}}{f(\theta_{3})^{2}
\sqrt{2\lambda E\cos^{2}\theta_{1}
+\frac{4E^{3/2}}{\lambda}g(\theta_{3})\sin^{3}\theta_{1}}},
\end{eqnarray}
where
\begin{eqnarray}
g(\theta) & = & \frac{\cos^{6}\theta +\sin^{6}\theta
+ k(k-2)\cos^{2}\theta\sin^{2}\theta}{f(\theta)^{6}}.
\end{eqnarray}


\end{document}